\documentclass[conference]{IEEEtran}
\twocolumn
\usepackage{cite} 
\usepackage{amsmath}%
\usepackage{amsfonts}%
\usepackage{amssymb}%
\usepackage{graphicx}
\newtheorem{theorem}{Theorem}

\usepackage{pdflscape}
\usepackage{multirow}

\newcommand{\Y}{\boldsymbol{Y}}

\newcommand{\SNR}{\textsf{SNR}}

\usepackage{pgf}
\usepackage{tikz}
\usetikzlibrary{arrows,automata}
\usepackage[latin1]{inputenc}
\usepackage{verbatim}

\begin{document}

\title{Phase Modulation for Discrete-time Wiener Phase \\ Noise Channels with Oversampling at High SNR}

\author{
\IEEEauthorblockN{Hassan Ghozlan}
\IEEEauthorblockA{Department of Electrical Engineering\\
University of Southern California\\
Los Angeles, CA 90089 USA\\
ghozlan@usc.edu}
\and
\IEEEauthorblockN{Gerhard Kramer}
\IEEEauthorblockA{
Institute for Communications Engineering \\
Technische Universit\"{a}t M\"{u}nchen \\
80333 Munich, Germany \\
gerhard.kramer@tum.de}
}

\maketitle

\begin{abstract}
A discrete-time Wiener phase noise channel model is introduced in which multiple samples are available at the output for every input symbol.
A lower bound on the capacity is developed.
At high signal-to-noise ratio (SNR), if the number of samples per symbol
grows with the square root of the SNR, 
the capacity pre-log is at least 3/4.
This is strictly greater than the capacity pre-log of the Wiener phase noise channel with only one sample per symbol, 
which is 1/2.
It is shown that amplitude modulation achieves a pre-log of 1/2
while phase modulation achieves a pre-log of at least 1/4.
\end{abstract}

\section{Introduction}
Communication systems often suffer from phase noise due to the instability of oscillators \cite{Demir2000}.
The characteristics of the phase noise process vary by application.
In systems with phase tracking devices, such as phase-locked loops (PLL), 
the residual phase noise follows a Tikhonov distribution \cite{ViterbiPLL1963}.
In Digital Video Broadcasting DVB-S2, an example of a satellite communication system, 
the phase noise process is modeled by the sum of the outputs of two infinite-impulse response filters 
driven by the same white Gaussian noise process \cite{CasiniDVBS2PN2004}.
In fiber-optic communication, the phase noise in laser oscillators 
is modeled by a Wiener process \cite{TkachDFB1986}.

For \emph{discrete-time} phase noise channels with a
stationary and ergodic phase noise process (whose entropy rate is finite),
Lapidoth showed that the capacity grows logarithmically with the signal-to-noise ratio (SNR) 
with a pre-log factor equal to 1/2 at high SNR \cite{LapidothPhaseNoise2002}. 
The two cases of Wiener phase noise and auto-regressive-moving-average (ARMA) phase noise fall into this class.
At finite SNR, numerical methods exist for computing (bounds on) the information rate 
for Wiener and ARMA phase noise
\cite{Dauwels2008, Barbieri2011, BarlettaLB, BarlettaUB, Barletta2012KalmanLB, LucaARMA2013}.

In \cite{GhozlanISIT2013, GhozlanGLOBECOM2013, LucaWhiteISIT2014, LucaWhiteCROWN2014}, \emph{continuous-time} phase noise channels are studied.
Continuous-time white phase noise is considered in \cite{LucaWhiteISIT2014, LucaWhiteCROWN2014}.
In \cite{GhozlanISIT2013} and \cite{GhozlanGLOBECOM2013}, a discrete-time phase noise channel is developed by discretizing a continuous-time Wiener phase noise channel 
by oversampling the output of an integrate-and-dump filter at the receiver.
It was shown in \cite{GhozlanISIT2013} that, at high SNR, 
the information rate grows logarithmically with SNR with a pre-log factor equal to 1/2
when the number of samples per symbol grows with the square root of the SNR. 
This result was established by employing amplitude modulation only.
It was shown in \cite{GhozlanGLOBECOM2013} through numerical simulations that oversampling improves the information rate
for Phase Shift Keying (PSK) modulation (see Fig. 5 in \cite{GhozlanGLOBECOM2013}).
The question of whether phase modulation can increase the pre-log factor at high SNR is left open.

We study in this paper a discrete-time channel model similar to \cite{LapidothPhaseNoise2002, Dauwels2008, Barbieri2011, BarlettaLB, BarlettaUB, Barletta2012KalmanLB, LucaARMA2013}, 
namely one without amplitude noise that would arise due to filtering before sampling \cite{GhozlanISIT2013, GhozlanGLOBECOM2013, LucaWhiteISIT2014, LucaWhiteCROWN2014}.
We do this as a first step towards addressing the more complex continuous-time model.
Our approach is similar to \cite{GhozlanISIT2013, GhozlanGLOBECOM2013} in that we consider oversampling receivers, 
where the oversampling rate increases with the square root of the SNR to achieve the maximum pre-log of 1/2 for amplitude modulation. 
However, as we will show, we achieve an additional pre-log of 1/4 by using only 2 samples per symbol. 

The paper is organized as follows. 
In Section \ref{sec:dt-model}, the discrete-time model of \cite{GhozlanISIT2013}
for the Wiener phase noise channel with oversampling is described 
and a simplified discrete-time channel model is introduced.
A lower bound on capacity of the simplified channel is derived in Section \ref{sec:capacity_lowerbound}
and the paper is concluded with Section \ref{sec:conc}.


\section{Discrete-time Model}
\label{sec:dt-model}
We use the following notation: 
$j=\sqrt{-1}$ , 
$^*$ denotes the complex conjugate, 
$\delta_D$ is the Dirac delta function, 
$\lceil \cdot \rceil$ is the ceiling operator.
We use $X^k$ to denote $(X_1,X_2,\ldots,X_k)$.
We describe the discrete-time model developed in \cite{GhozlanISIT2013}.
Let $X^n$ be the input symbols. For every input symbol, there are $L$ output samples.
The $k$-th output sample is
\begin{align}
Y_k^{\textsf{full}} = X_{\lceil k/L \rceil} \Delta \ e^{j \Theta_k} \ F_k + N_k
\label{eq:Yk}
\end{align}
where $k=1,\ldots,n L$ and $\Delta = 1/L$.
The process $\{N_k\}$ is an independent and identically distributed (i.i.d.) circularly-symmetric complex Gaussian process with mean $0$ and 
$\mathbb{E}[ |N_k|^2 ] = \sigma^2_N \Delta$
while the process $\{\Theta_k\}$ is the discrete-time Wiener process
\begin{align}
	\Theta_{k+1} = \Theta_{k} + W_{k}
\end{align}
where
$\Theta_1$ is uniform on $[-\pi,\pi)$ and
$\{W_k\}$ is an i.i.d. real Gaussian process with mean $0$ and $\mathbb{E}[ |W_k|^2 ] = \sigma^2_W = 2\pi \beta \Delta$.
Moreover, $\{W_k\}$ is independent of $\{N_k\}$.
The random variable $F_k$ is defined as
\begin{align}
F_k \equiv \frac{1}{\Delta} \int_{(k-1) \Delta}^{k \Delta} e^{j(\Theta(\tau)-\Theta((k-1) \Delta))} \ d\tau
\label{eq:Fk_def}
\end{align}
and $\Theta(t)$ is a continuous-time Wiener process:
\begin{align}
  \Theta(t) = \Theta(0) + \int_0^t W(\tau) d\tau
\label{eq:Thetat}
\end{align}
where $\Theta(0)$ is uniform  on $[-\pi,\pi)$ and 
$W(t)$ is a real Gaussian process with
\begin{align}
&\mathbb{E}\left[ W(t) \right] = 0 \\
&\mathbb{E}\left[ W(t_1) W(t_2)\right] = 2\pi \beta ~ \delta_D(t_2-t_1) .
\label{eq:Wt_statistics}
\end{align}
The parameter $\beta>0$ is the full-width at half-maximum (FWHM)
of the power spectral density of $e^{j\Theta(t)}$.
A power constraint is imposed on the transmitted symbols
\begin{align}
	\frac{1}{n} \sum_{m=1}^{n} \mathbb{E}[|X_m|^2] \leq P.
	\label{eq:dt_power_constraint}
\end{align}
The signal-to-noise ratio $\SNR$ is defined as 
$\SNR = {P}/{\sigma^2_N}$.

The model we adopt in this paper does not include the effect of filtering modeled by $\{F_k\}$. 
More specifically, the $k$-th output of the simplified model is
\begin{align}
Y_k = X_{\lceil k/L \rceil} \Delta \ e^{j \Theta_k} + N_k
\label{eq:Yk_approx}
\end{align}
where $\{\Theta_k\}$ and $\{N_k\}$ are the same processes defined earlier.
We remark that the process $\{\Theta_k\}$ is not stationary 
but $\{e^{j \Theta_k}\}$ is stationary.

\section{Lower Bound on Capacity}
\label{sec:capacity_lowerbound}
Define
$\Y_k \equiv (Y_{(k-1) L + 1},Y_{(k-1) L + 2},\ldots,Y_{(k-1) L + L})$.
We also use $\Y^k$ as a shorthand for $(\Y_1,\Y_2,\ldots,\Y_k)$.
The capacity of (\ref{eq:Yk_approx}) is given by
\begin{align}
C(\SNR) = \lim_{n \rightarrow \infty} \frac{1}{n} \sup I(X^n;\Y^n)
\label{eq:capacity-def}
\end{align}
where the supremum is over all of possible joint distributions 
of the input symbols satisfying the power constraint.
For a given input distribution, the achievable rate $R$ is given by
\begin{align}
R(\SNR) = I(X;\Y) \equiv \lim_{n \rightarrow \infty} \frac{1}{n} I(X^n;\Y^n).
\label{eq:I_X_Y-def}
\end{align}

Our main result is the following theorem.
\begin{theorem}
If $L = \lceil \beta \sqrt{\SNR} \rceil$ and
the input $X^n$ is i.i.d. with $\arg(X_k)$ independent of $|X_k|$ for $k=1,\ldots,n$ such that 
$\arg(X_k)$ is uniformly distributed over $[-\pi,\pi)$ and
$(|X_k|^2-P/2)$ is exponentially distributed with mean $P/2$,
then
\begin{align}
\lim_{\SNR \rightarrow \infty} I(X;\Y) - \frac{3}{4} \log{\SNR}
\geq \text{constant}.
\label{eq:inforate-overall}
\end{align}

\end{theorem}
As a corollary, the capacity pre-log satisfies
\begin{align}
\lim_{\SNR \rightarrow \infty} \frac{C(\SNR)}{\log{\SNR}} \geq \frac{3}{4}.
\label{eq:prelog}
\end{align}

We outline the proof in the rest of this section.
Without loss of generality, let $\sigma_N^2=1$.
Define
$X_A \equiv |X|$ and $\Phi_X \equiv \angle X$.
We decompose the mutual information using the chain rule into two parts:
\begin{align}
I(X^n;\Y^n) 
&= I(X_{A}^n;\Y^n) + I(\Phi_{X}^n;\Y^n|X_{A}^n).
\label{eq:I_X1n_Y1n}
\end{align}
The first term represents the contribution of the amplitude modulation while 
the second term represents the contribution of the phase modulation.
First, we analyze the amplitude modulation term.
We have
\begin{align}
I(X_{A}^n;\Y^n)
&\stackrel{(a)}{=} \sum_{k=1}^n I(X_{A,k};\Y^n|X_{A}^{k-1}) \nonumber \\
&\stackrel{(b)}{\geq} \sum_{k=1}^n I(X_{A,k};V_k| X_{A}^{k-1})
\label{eq:I_XA1n_Y1n_LB}
\end{align}
where $V_k$ is a deterministic function of $(\Y^n,X_{A}^{k-1})$.
Step
$(a)$ follows from the chain rule of mutual information and
$(b)$ follows from the data processing inequality.
We choose
\begin{align}
V_k = \sum_{\ell=1}^L |Y_{(k-1)L+\ell}|^2.
\label{eq:V_def}
\end{align}
When $X^n$ is i.i.d., the pair $(X_{A,k},V_k)$ with $V_k$ defined in (\ref{eq:V_def}) is independent of $X_{A}^{k-1}$ and therefore
\begin{align}
I(X_{A,k};V_k| X_{A}^{k-1}) = I(X_{A,k};V_k).
\end{align}
By using the auxiliary-channel lower bound theorem in \cite[Sec. VI]{Arnold2006}, we have
\begin{align}
I(X_{A,k};V_k)
\geq \mathbb{E}[\log{Q_{V|X_A}(V_k|X_{A,k})}] - \mathbb{E}[\log{Q_{V,k}(V_k)}]
\label{eq:I_XA_V_LB_AuxCh}
\end{align}
where $Q_{V|X_A}(v|x_A)$ is an arbitrary auxiliary channel and
\begin{align}
Q_{V_k}(v) \equiv \int p_{X_{A,k}}(x_A) Q_{V|X_A}(v|x_A) dx_A
\label{eq:QV_def}
\end{align}
where $p_{X_{A,k}}(\cdot)$ is the \emph{true} distribution of $X_{A,k}$,
i.e.,
$Q_V(\cdot)$ is the output distribution obtained by connecting the true input source to the auxiliary channel.
We choose the auxiliary channel
\begin{align}
Q_{V|X_A}(v|x_A) 
= \frac{1}{\sqrt{4 \pi x_A^2 \Delta^2 \sigma^2_N}}
\exp\left(- \frac{(v-x_A^2 \Delta - \sigma^2_N)^2}{4 x_A^2 \Delta^2 \sigma^2_N} \right).
\label{eq:Q_V|XA}
\end{align}
Following steps similar to those in \cite{GhozlanISIT2013}, it can be shown that
if $X_{A}^n$ is i.i.d. with $|X_k|^2$ distributed according to $p_{X_P}$ 
for $k=1,\ldots,n$ where
\begin{align}
p_{X_P}(|x|^2) = \left\{ 
  \begin{array}{ll}
  \frac{2}{P} \exp\left(1-\frac{2|x|^2}{P}\right),	& |x|^2 \geq P/2 \\
  0,			& \text{otherwise}
  \end{array}
 \right.
\label{eq:P_XP}
\end{align}
then
\begin{align}
\lim_{\SNR \rightarrow \infty} I(X_{A};\Y) - \frac{1}{2} \log{\SNR}
\geq - 2  - \frac{1}{2} \log(8 \pi)
\label{eq:inforate-amplitude-contrib}
\end{align}
where
\begin{align}
I(X_{A};\Y) \equiv \lim_{n \rightarrow \infty} \frac{1}{n} I(X_{A}^n;\Y^n).
\label{eq:I_XA_Y-def}
\end{align}

Next, we turn our attention to the contribution of the phase modulation.
By using the chain rule, we have
\begin{align}
I(\Phi_{X}^n;\Y^n|X_{A}^n)
&= \sum_{k=1}^n I(\Phi_{X,k};\Y^n|X_{A}^n,\Phi_{X}^{k-1}) \nonumber \\
&\stackrel{(a)}{\geq} \sum_{k=2}^n I(\Phi_{X,k};\Y^n|X_{A}^n,\Phi_{X}^{k-1}) \nonumber \\
&\stackrel{(b)}{\geq} \sum_{k=2}^n I(\Phi_{X,k};\tilde{Y}_k |X_{A}^n,\Phi_{X}^{k-1})
\label{eq:I_PhiX1n_Y1n|XA1n_LB1}
\end{align}
where
$\tilde{Y}_k$ is a deterministic function of $(\Y^n,X_{A}^n,\Phi_{X}^{k-1})$.
Inequality $(a)$ follows from the non-negativity of mutual information
and $(b)$ follows from the data processing inequality.

At high SNR, we use some intuition to choose a reasonable processing of
$(\Y^n,X_{A}^n,\Phi_{X}^{k-1})$ for decoding $\Phi_{X,k}$:
\begin{enumerate}
\item
Since only the past inputs $X^{k-1}$ are available, 
the future outputs $\Y_{k+1}^n$ are not very useful for estimating $\Theta_{k-1}$.

\item
Since $\{\Theta_k\}$ is a first-order Markov process, 
the most recent past input symbol $X_{k-1}$ and the most recent output sample $Y_{(k-1)L}$ are the most useful for estimating $\Theta_{k-1}$.
A simple estimator is
\begin{align}
e^{j \widehat{\Theta}_{k-1}}
\equiv \frac{Y_{(k-1)L}}{X_{k-1} \Delta} 
&= e^{j \Theta_{(k-1)L}} + \frac{N_{(k-1)L}}{X_{k-1} \Delta} \nonumber\\
&= e^{j \Theta_{(k-1)L}}  \left( 1 + \tilde{Z}_{k-1}^*  \right)
\label{eq:Theta_hat}
\end{align}
where
\begin{align}
\tilde{Z}_k \equiv \frac{N_{kL}^* \ e^{-j \Theta_{kL}} }{X_k^* \Delta}.
\label{eq:Zk_tilde_def}
\end{align}

\item
Given the current input amplitude $|X_k|$ and the estimate of $\Theta_{k-1}$, 
the first sample $Y_{(k-1)L+1}$ in $\Y_k$ is the most useful for decoding $\Phi_{X,k}$
because the following samples become increasingly corrupted by the phase noise.
We scale $Y_{(k-1)L+1}$ to normalize the variance of the additive noise and write
\begin{align}
\frac{Y_{(k-1)L+1}}{\sqrt{\Delta}}
= \bigg( |X_k| \sqrt{\Delta} e^{j\Phi_{X,k}} + \tilde{N}_k \bigg) e^{j \Theta_{(k-1)L+1}}
\label{eq:Y_normalized}
\end{align}
where 
\begin{align}
\tilde{N}_{k} \equiv \frac{N_{(k-1)L+1} \ e^{-j \Theta_{(k-1)L+1}} }{\sqrt{\Delta}}.
\label{eq:Nk_tilde_def}
\end{align}

\end{enumerate}
To summarize, we choose
\begin{align}
\tilde{Y}_k
= \frac{Y_{(k-1)L+1}}{\sqrt{\Delta}} \left( \frac{Y_{(k-1)L}}{X_{k-1} \Delta}  \right)^*.
\label{eq:Yk_tilde_def}
\end{align}
It follows from (\ref{eq:Yk_tilde_def}), (\ref{eq:Theta_hat}) and (\ref{eq:Y_normalized}) that
\begin{align}
\tilde{Y}_k
= \bigg( |X_k| \sqrt{\Delta} e^{j\Phi_{X,k}} + \tilde{N}_k \bigg) \left( 1 + \tilde{Z}_{k-1} \right) e^{j W_{(k-1)L+1}}
\label{eq:Yk_tilde}
\end{align}
where $\tilde{N}_k$ and $\tilde{Z}_{k-1}$ are statistically independent and
\begin{align}
&\tilde{N}_k \sim \mathcal{N}_{\mathbb{C}}(0,1) \\
&\tilde{Z}_{k-1} \Big| \{|X_{k-1}|=|x_{k-1}|\} \sim \mathcal{N}_{\mathbb{C}}\left(0,\frac{1}{|x_{k-1}|^2 \Delta}\right)
\end{align}
which means that,
conditioned on $\{|X_{k-1}|=|x_{k-1}|\}$, $\tilde{Z}_{k-1}$ is a Gaussian random variable 
with mean $0$ and variance $1/(|x_{k-1}|^2 \Delta)$.
Moreover, $W_{(k-1)L+1}$ is statistically independent of $\tilde{N}_k$ and $\tilde{Z}_{k-1}$.
The choice of $\tilde{Y}_k$ in (\ref{eq:Yk_tilde_def}) implies that
\begin{align}
I(\Phi_{X,k};\tilde{Y}_k |X_{A}^n,X^{k-1})
&= I(\Phi_{X,k};\tilde{Y}_k |X_{A,k},X_{k-1}).
\label{eq:I_PhiX_Ytilde_LEBOWSKI}
\end{align}
Define $\tilde{\Phi}_{Y,k} \equiv \angle{\tilde{Y}_k}$ and
\begin{align}
&Q_{\tilde{\Phi}_{Y}|\Phi_{X}} \left(\phi_y \big| \phi_x \right) 
\equiv \frac{\exp(\alpha \cos(\phi_y-\phi_x) )}{2 \pi I_0(\alpha)}.
\label{eq:q_PhiY|PhiX_new}
\end{align}
where $I_0(\cdot)$ is the zeroth-order modified Bessel function of the first kind and $\alpha > 0$.
This distribution is known as Tikhonov (or von Mises) distribution \cite{Mardia1972}.
Furthermore, define
\begin{align}
&Q_{\tilde{\Phi}_{Y,k}|X_{A,k},X_{k-1}}\left(\phi_y \big| |x_k|,x_{k-1} \right) \nonumber\\&
\equiv \int_{-\pi}^{\pi}
p_{\Phi_{X,k}|X_{A,k},X_{k-1}} \left(\phi_x \big| |x_k|,x_{k-1} \right) 
Q_{\tilde{\Phi}_{Y}|\Phi_{X}} (\phi_y | \phi_x)
d\phi_x \nonumber\\&
= \frac{1}{2 \pi}.
\label{eq:q_PhiY_new}
\end{align}
The last equality holds because $X_1,\ldots,X_n$ are statistically independent
and $\Phi_{X,k}$ is independent of $X_{A,k}$ with a uniform distribution on $[-\pi,\pi)$.
We have
\begin{align}
&I(\Phi_{X,k};\tilde{Y}_k |X_{A,k},X_{k-1}) \nonumber\\
&\stackrel{(a)}{\geq} I(\Phi_{X,k};\tilde{\Phi}_{Y,k} |X_{A,k},X_{k-1}) \nonumber\\
&\stackrel{(b)}{\geq} 
\mathbb{E}\left[\log Q_{\tilde{\Phi}_{Y}|\Phi_{X}} (\tilde{\Phi}_{Y,k}|\Phi_{X,k}) \right]  \nonumber\\&~ - 
\mathbb{E}\left[\log Q_{\tilde{\Phi}_{Y,k}|X_{A,k},X_{k-1}} \Big(\tilde{\Phi}_{Y,k} \big| |X_k|,X_{k-1} \Big) \right] \nonumber\\
&\stackrel{(c)}{=} \log(2\pi) - \log( 2 \pi I_0(\alpha) ) + \alpha \mathbb{E}\left[ \cos(\tilde{\Phi}_{Y,k}-\Phi_{X,k}) \right] \nonumber\\
&= - \log( I_0(\alpha) ) + \alpha \mathbb{E}\left[ \cos(\tilde{\Phi}_{Y,k}-\Phi_{X,k}) \right] \nonumber\\
&\stackrel{(d)}{\geq} \frac{1}{2} \log\alpha - \alpha 
    + \alpha \mathbb{E}\left[ \cos(\tilde{\Phi}_{Y,k}-\Phi_{X,k}) \right]
\label{eq:I_PhiX_PhiY_LB1} \\
&\geq \frac{1}{2} \log{\alpha} - \alpha \frac{\sigma^2_{W}}{2} - \frac{4 \alpha}{\SNR \Delta}
\label{eq:I_PhiX_Ytilde_LB_LEBOWSKI}
\end{align}
where
$(a)$ follows from the data processing inequality,
$(b)$ follows by extending the result of the auxiliary-channel lower bound theorem in \cite[Sec. VI]{Arnold2006},
$(c)$ follows from (\ref{eq:q_PhiY|PhiX_new}) and (\ref{eq:q_PhiY_new}),
$(d)$ follows from \cite[Lemma 2]{FOCUS2010}
\begin{align}
I_0(z) \leq \frac{\sqrt{\pi}}{2} \frac{e^z}{\sqrt{z}} \leq \frac{e^z}{\sqrt{z}}
\label{eq:BesselI0_UB}
\end{align}
and $(d)$ holds because\footnote{The proof is omitted.}
\begin{align}
\mathbb{E}\left[\cos(\tilde{\Phi}_{Y,k}-\Phi_{X,k})\right] 
\geq 1 - \frac{\sigma^2_{W}}{2} - \frac{4}{\SNR \Delta}
\label{eq:Ecos_PhiYtilde_minus_PhiX_LB}
\end{align}
for $\SNR \Delta > 2$.
It follows from (\ref{eq:I_PhiX1n_Y1n|XA1n_LB1}), (\ref{eq:I_PhiX_Ytilde_LEBOWSKI}) and (\ref{eq:I_PhiX_Ytilde_LB_LEBOWSKI}) that
\begin{align}
\frac{1}{n} I(\Phi_{X}^n;\Y^n|X_{A}^n)
&\geq \frac{n-1}{n}
\left[ \frac{1}{2} \log{\alpha} - \alpha \pi\beta\Delta - \frac{4 \alpha}{\SNR \Delta} \right].
\label{eq:I_PhiX1n_Y1n|XA1n_LB2}
\end{align}
Hence, we have
\begin{align}
I(\Phi_{X};\Y|X_{A})
&\equiv \lim_{n \rightarrow \infty} \frac{1}{n} I(\Phi_{X}^n;\Y^n|X_{A}^n) \label{eq:I_PhiX_Y|XA-def} \\
&\geq \frac{1}{2} \log{\alpha} - \alpha \pi\beta\Delta - \frac{4 \alpha}{\SNR \Delta}.
\label{eq:I_PhiX_Y|XA_LB_LEBOWSKI}
\end{align}
Suppose $L$
grows with $\SNR$ such that 
\begin{align}
 L = \left\lceil \beta \sqrt{\SNR} \right\rceil.
\end{align}
Since $\Delta = 1/L$, we have
\begin{align}
\lim_{\SNR \rightarrow \infty} \SNR \Delta^2 = \frac{1}{\beta^2}.
\end{align}
Therefore, by setting $\alpha = \SNR \Delta$ and taking the limit of $\SNR$ tending to infinity, we have
\begin{align}
\lim_{\SNR \rightarrow \infty} I(\Phi_{X};\Y|X_{A}) - \frac{1}{4} \log{\SNR}
\geq \log{\frac{1}{\beta}}-\frac{\pi}{\beta}-4.
\label{eq:inforate-phase-contrib}
\end{align}
The last equation implies that the phase modulation contributes $1/4$ to the pre-log of the information rate 
when oversampling is employed.
It follows from (\ref{eq:I_X_Y-def}), (\ref{eq:I_XA_Y-def}), (\ref{eq:I_PhiX_Y|XA-def}) and (\ref{eq:I_X1n_Y1n}) that
\begin{align}
I(X;\Y) = I(X_A;\Y) + I(\Phi_{X};\Y|X_{A})
\label{eq:I_X_Y}
\end{align}
Combining (\ref{eq:inforate-amplitude-contrib}) and (\ref{eq:inforate-phase-contrib}) yields (\ref{eq:inforate-overall}).

It is worth pointing out that the phase modulation pre-log of 1/4 requires only 2 samples per symbol 
for which the time resolution, $1/\Delta$, grows as the square root of the SNR. 
It is interesting to contemplate whether another receiver, e.g., a non-coherent receiver, 
can achieve the maximum amplitude modulation pre-log of 1/2 but requires only 1 sample per symbol. 
If so, one would need only 3 samples per symbol to achieve a pre-log of 3/4.

\section{Conclusion}
\label{sec:conc}
We studied a discrete-time model of a Wiener phase noise channel with oversampling. 
We showed that, at high SNR, the capacity grows logarithmically with SNR with a pre-log of at least 3/4 
if the number of samples per symbol grows with the square root of the SNR.
It was found that amplitude modulation and phase modulation can achieve pre-log factors of 1/2 and 1/4, respectively.
In fact, the phase modulation pre-log of 1/4 requires only 2 samples per symbol.

\section*{Acknowledgment}
H. Ghozlan was supported by a USC Annenberg Fellowship and NSF Grant CCF-09-05235.
G. Kramer was supported by an Alexander von Humboldt Professorship endowed by
the German Federal Ministry of Education and Research.

\bibliographystyle{unsrt}
\bibliography{ref13-short}

\end{document}